\documentstyle[prd,aps,floats,graphicx]{revtex}

\begin{document}
\preprint{astro-ph/yymmxxx}
\preprint{SUSX-TH/02-008}
\draft

\renewcommand{\topfraction}{0.99}
\renewcommand{\bottomfraction}{0.99}

\twocolumn[\hsize\textwidth\columnwidth\hsize\csname
@twocolumnfalse\endcsname

\title{Tracking quintessential inflation from brane worlds}
\author{N. J. Nunes and E. J. Copeland}
\address{Centre for Theoretical Physics, CPES, University of Sussex,
Brighton BN1 9QJ, United Kingdom}
\date{\today}
\maketitle
\begin{abstract}
We analyse the mechanism of quintessential inflation in brane world
scenarios for a number of particle physics inspired scalar
potentials. We constrain the parameter space of those scalar
potentials and comment on the likelihood that we could discriminate these
models from standard inflation, based on upcoming large scale
structure and cosmic microwave background observations.
\end{abstract}


\pacs{PACS numbers: 98.80.Cq \hfill
SUSX-TH/02-008 \hspace{1cm} astro-ph/0204115}
\vskip2pc]

\section{Introduction}
There has been interest, recently, on the
cosmological implications arising in
a certain class of brane world scenarios where 
the Friedmann equation is modified at very high energies
\cite{bin}.
The key feature is that the modification
makes it easier to obtain inflation in the early universe, by
contributing extra friction to the scalar field equation of motion
\cite{quadref,maart,maeda}. Analysing this situation, it has been
explicitly demonstrated how inflation naturally occurs in
such models even when the potential is too steep to allow conventional
inflation to occur \cite{copeland1,lidsey}. Inflation ends when the brane world corrections
begin to lose their dominance, and reheating takes place as a result of gravitational
particle production, rather than the usual inflaton decay mechanism. In this case
the inflaton
energy density subsequently redshifts sufficiently quickly that the produced
radiation comes to dominate. One of the
nice features this allows for is that the inflaton potential does not have
to have a feature in it, and the field does not have to decay
completely as we do not require a minimum to the potential about which the
inflaton normally oscillates as it decays. Instead, we can consider
the intriguing
possibility that the inflaton actually survives as inflation ends. As the
universe evolves through radiation to matter domination the inflaton 
potential does
not play much of a role, however, at a recent redshift of $z \sim 1$, it 
once again comes to dominate driving the universe to an accelerated
expansion. This mechanism has been named
``quintessential inflation'' scenario \cite{PV,quinfl}.
In this paper we investigate this possibility in more depth.
Similar approaches have been
adopted in the brane world context \cite{huey,majumdar,sahni}
although in this paper we will be analysing a broader class of potentials
and discussing the likelihood of being able to detect the 
presence of such a field in the next generation of cosmic microwave 
background (CMB)
observations. The class of potentials we will investigate 
include inverse power law, exponential and  potentials motivated by
supergravity models.

We consider the five--dimensional brane scenario, in which the
Friedmann equation
is modified from its usual form, becoming \cite{bin}
\begin{equation}
H^2 = \frac{\kappa^2}{3} ~\rho ~
\left[ 1 + \frac{\rho}{2\lambda_{\rm b}} \right]
    + \frac{\Lambda_4}{3} + \frac{{\cal E}}{a^4} \,,
\end{equation}
where $\kappa^2 \equiv 8 \pi G = 8 \pi / M^2_4$ and $M_4$
is the four-dimensional Planck mass.
$\Lambda_4$ is the
four-dimensional cosmological constant and the final term
represents the influence of bulk gravitons on the brane. 

The brane tension
$\lambda_{\rm b}$ relates the four and five-dimensional Planck masses via
\begin{equation}
M_4 = \sqrt{\frac{3}{4\pi}} \,
\left( \frac{M_5^2}{\sqrt{\lambda_{\rm b}}} \right) \, M_5 \,,
\end{equation}
and is constrained by the requirement of successful nucleosynthesis as $\lambda_{\rm b}
> (1 \, {\rm MeV})^4$. We assume that the
four--dimensional cosmological constant cancels out by some (as yet
undiscovered) mechanism, and once inflation begins the final term will rapidly
become unimportant, leaving us with
\begin{equation}
\label{Friedmannmodify}
H^2 = \frac{\kappa^2}{3} \, \rho \, \left[ 1 + \frac{\rho}{2\lambda_{\rm b}}
    \right] \,.
\end{equation}
We assume that the scalar field is confined to the brane, so that its field
equation has the standard form
\begin{equation}
\label{eqom}
\ddot{\phi} + 3 H \dot{\phi} + \frac{dV}{d\phi} = 0 \,.
\end{equation}

Following Ref. \cite{ford} the density of particles produced after
inflation is
\begin{equation}
\label{rhoend}
\rho_{\rm R}^{\rm end} = 0.01 g_{\rm p} H_{\rm end}^4 \,,
\end{equation}
where $g_{\rm p}$ is the number of scalar fields involved in particle
production, likely to be $ 10 \lesssim g_{\rm p} \lesssim 100$, 
and $H_{\rm end}$ is 
the value of the Hubble parameter at the end of inflation.

In our brane world scenario, the ratio of the energy densities
of the scalar field to the relativistic particles is well determined
from Eq.~(\ref{rhoend}),
\begin{equation}
\label{ratio}
\frac{\rho_{\phi}^{\rm end}}{\rho_{\rm R}^{\rm end}} \approx  
 \frac{ 36 \lambda_{\rm b}^2}
{0.01 g_{\rm p} \kappa^4 V_{\rm end}^3}  \,.
\end{equation}
Hence, the ratio of the energy densities is inversely proportional to
the number of fields involved in gravitational production $g_{\rm
p}$. Moreover, it decreases as the time between
the end of inflation $a_{\rm end}$ 
and the maximum
of the energy density of the produced relativistic particles
$a_{\rm p}$ increases.
Defining $\Delta \equiv \log (a_{\rm p}/a_{\rm end})$, we find from
our numerical simulations that this value is weakly dependent on 
the parameters of
the scalar potential or the number of $e$--foldings of inflation, and
spans the interval $0.29 \lesssim \Delta \lesssim 0.4$.
Therefore,  when constraining the parameter space of
scalar potentials, we will consider both the average 
case of $\Delta = 0.35$ with
$g_{\rm p} = 100$, and the case of instant reheating 
$\Delta = 0$ with $g_{\rm p} = 10$, which corresponds to the situation
when the quoted ratio is its maximum.

\section{Inflationary parameters}

It proves useful to define the
slow--roll parameters, analogously to the usual inflationary
case. Following Refs. \cite{maart,huey}  we define
$\epsilon$, $\eta$ and $\xi^2$, generalising the usual ones
\cite{LL}, by
\begin{eqnarray}
\label{epsilon}
\epsilon & \equiv & \frac{1}{2\kappa^2}~\left( \frac{V'}{V}
\right)^2~ \frac{1+V/\lambda_{\rm b}}{(1+V/2\lambda_{\rm b})^2} \,,
\\
\eta & \equiv & \frac{1}{\kappa^2}~ \frac{V''}{V}~
\frac{1}{1+V/2\lambda_{\rm b}} \,, \\
\xi^2 & \equiv & \frac{1}{\kappa^4}~ \frac{V' V'''}{V^2}~
\frac{1}{(1+V/2\lambda_{\rm b})^2} \,
\end{eqnarray}
where prime indicates a $\phi$--derivative and the slow-roll
approximation has been employed.

As the Universe inflates, the associated number of $e$--foldings 
of the scale factor $a$, 
is given by
\begin{equation}
\label{efoldings}
N \equiv \int_a^{a_{\rm end}} d \ln a
\simeq - \kappa^2 \int_\phi^{\phi_{{\rm end}}} \frac{V}{V'} \,
\left(1 +
\frac{V}{2\lambda_{\rm b}} \right) \, d\phi \,,
\end{equation}
where $a_{\rm end}$ corresponds to the end of inflation, 
given by the condition $\epsilon =1$. 
In this model, the number of $e$--foldings corresponding to scales 
entering the Hubble radius today can be unambiguously determined
\cite{sahni}. The scale that leaves the Hubble radius
during inflation and re--enters today is $k = a_N H_N = a_0 H_0$, i.e.
\begin{equation}
1 = \frac{k}{a_0 H_0} = \frac{a_N}{a_{\rm end}}
\frac{a_{\rm end}}{a_{\rm p}}
\frac{a_{\rm p}}{a_0} \frac{H_N}{H_0} \,.
\end{equation}
The subscript $N$ corresponds to the value of the quantity $N$
$e$--foldings from the end of inflation.
Noting that $a_N/a_{\rm end} = \exp(-N)$ and 
$\rho_{\rm R}^{\rm end} = \rho_{\rm R}^0~(a_0/a_{\rm p})^4$ we find
\begin{equation}
\label{efolds1}
N = \ln \left[ \left( \frac{3 \Omega_{\rm R}}{0.01 g_{\rm p}}
\right)^{1/4}
(\kappa H_0)^{-1/2} \frac{a_{\rm end}}{a_{\rm p}} 
\frac{H_{\rm N}}{H_{\rm end}}  \right] \,.
\end{equation}
where $\Omega_{\rm R} =2.471 h^{-2} \times 10^{-5}$ is the 
fractional energy density in radiation today.

The amplitude of the scalar and tensor perturbations produced during
inflation are given by \cite{maart,huey}
\begin{equation}
A_{\rm S}^2 = \frac{\kappa^4}{150 \pi^2} \, \frac{V}{\epsilon} \,
    \left( 1+\frac{V}{2\lambda_{\rm b}} \right) \, \left( 1+\frac{V}{\lambda_{\rm b}} \right)  \,,
\label{scalaramp}
\end{equation}

\begin{equation}
A_{\rm T}^2 = \frac{\kappa^4}{150 \pi^2} \, V \,
    \left( 1+\frac{V}{2\lambda_{\rm b}} \right) \, F^2(V/\lambda_{\rm b})  \,,
\label{tensoramp}
\end{equation}
where
\begin{equation}
F^2(x) \equiv \left[ \sqrt{1+x^2} -
x^2 \sinh^{-1}\left(\frac{1}{x}\right)\right]^{-1}
\end{equation}
and
\begin{equation}
x \equiv \left({3 \over 4 \pi \lambda_{\rm b}} \right)^{1/2} H M_4 =
\left[{2 V \over \lambda_{\rm b}} \left( 1+\frac{V}{2\lambda_{\rm b}} \right)
\right]^{1/2}.
\end{equation}

In the low energy limit, $\rho \ll \lambda_{\rm b} ~(x \ll 1)$,
$F^2 \approx 1$, whereas $F^2 \rightarrow 3V/2\lambda_{\rm b}$ in the
high energy limit.

Having obtained the amplitude of the scalar and tensor perturbations
the corresponding scalar and tensor spectral indices can be obtained. 
Using 

\begin{equation}
{d \over d \ln k} \approx - {V' \over 3 H^2} {d \over d \phi} \,,
\end{equation}
we find \cite{huey}
\begin{equation}
\label{ns}
n_{\rm S}-1 \equiv  \frac{d \ln A^2_{\rm S}}{d \ln k} = -6\epsilon + 2\eta \,,
\end{equation}
\begin{equation}
n_{\rm T} \equiv  \frac{d \ln A^2_{\rm T}}{d \ln k} = -2\epsilon
\left( 1+ 2~\frac{F'}{F}\frac{V}{V'}~
\frac{1+V/2\lambda_{\rm b}}{1+V/\lambda_{\rm b}} \right) \,.
\label{nt}
\end{equation}

If we define the quantity $\delta$ to be 0 for the standard
inflationary scenario (without the brane corrections) 
and $\delta = 1$ in the brane inspired case, then we can
simplify Eq.~(\ref{nt}) for the low and high energy limits as,
\begin{equation}
n_{\rm T} = -(2+\delta)~ \epsilon \,.
\end{equation}
We also obtain that the consistency equation is the same in either
limit \cite{huey},
\begin{equation}
n_{\rm T} = -2~ \frac{A_{\rm T}^2}{A_{\rm S}^2} \,.
\end{equation}

The derivatives of the scalar and tensor spectral indices give in the
limits of high and low energies
\begin{eqnarray}
\frac{d n_{\rm S}}{d \ln k} &=& -(24-6 \delta) \epsilon^2 +
16 \epsilon \eta - 2 \xi^2 \,, \\
\frac{d n_{\rm T}}{d \ln k} &=& -(8+\delta) \epsilon^2 +
(4+2 \delta) \epsilon \eta \,.
\end{eqnarray}

In the next section we turn our attention to a number of particle physics
inspired potentials, to determine how the brane corrections modify the 
dynamics
of the associated scalar field, leading to inflation in a number of 
scenarios where
inflation would not have normally been present. In the following
section we will
proceed to investigate the class of models where the inflaton field
then becomes the quintessence field by today.

\section{Brane Inflation}
\label{braneinflation}
\subsection{Exponential potentials.}
The simplest case to investigate involves the dynamics associated 
with a pure exponential potential \cite{copeland1},
\begin{equation}
V(\phi) = V_0 \exp (\alpha \kappa \phi) \,. \label{exppotential}
\end{equation}
For completeness we recall the key results.
During slow--roll inflation, from Eq.~(\ref{epsilon}) we have

\begin{equation}
\epsilon \simeq \frac{2\alpha^2\lambda_{\rm b}}{V} \,.
\end{equation}
Inflation ends when $\epsilon = 1$, giving
\begin{equation}
\label{Vend}
V_{{\rm end}} \simeq 2\alpha^2 \lambda_{\rm b} \,,
\end{equation}
which implies that typically the term quadratic in the
density still dominates the linear term at the end of inflation.

The potential $V_N$,  $N$ $e$--foldings from the end of
inflation can be obtained from Eq.~(\ref{efoldings}) 
and is given by the simple formula
\begin{equation}
\label{VN}
V_N = V_{{\rm end}} \left( N+1 \right) \,.
\end{equation}

The amplitude of density perturbations generated during the
inflationary period fixes the brane tension
\begin{equation}
\label{lambdab}
\lambda_{\rm b} = \left[ \frac{2 \kappa^4}{75 \pi^2} \alpha^6
(1+N)^4 \right]^{-1} A_{\rm S}^2 \,,
\end{equation}
which significantly is independent of the mass parameter $V_0$. 

From Eqs.~(\ref{Friedmannmodify}) and (\ref{VN}), 
we have, $H_N = (N+1) H_{\rm end}$; and
substituting into Eq.~(\ref{efolds1}) we find that $N = 70.7$ for $\Delta =
0.35$, $g_{\rm p} = 100$ and $N = 72.1$ for $\Delta =
0$, $g_{\rm p} = 10$.

Inserting this 
into Eq.~(\ref{lambdab}), and using the observed value from COBE of 
$A_{{\rm S}} = 2 \times 10^{-5}$ \cite{COBEnorm}, we find that 
\begin{equation}
\label{COBE2}
\lambda_{\rm b} \simeq \frac{1 \times 10^{-17}}{\alpha^6} \, M_4^4 =
    \left(\frac{10^{15} \, {\rm GeV}}{\alpha^{3/2}}\right)^4 \,,
\end{equation}
indicating a brane tension of order the Grand Unified Theories scale.

For the scalar and tensor spectral indices and their derivatives we find,
\begin{eqnarray}
\label{spectral1}
n_{\rm S} -1 &=& (1+N)~\frac{d n_{\rm S}}{d \ln k} = -\frac{4}{1+N} \,,
\\
n_{\rm T} &=& (1+N)~\frac{d n_{\rm T}}{d \ln k} = -\frac{3}{1+N} \,.
\end{eqnarray}

\subsection{Power law potentials}
For power law potentials,
\begin{equation}
\label{powerlawpot}
V(\phi) = V_0~ (\kappa \phi)^n \,,
\end{equation}
during slow--roll inflation we have
\begin{equation}
\epsilon \simeq \frac{2 n^2 \lambda_{\rm b}}{V_0}~
                (\kappa \phi)^{-n-2} \,.
\end{equation}
For the particular case of $n = -2$ the condition for inflation is 
$\lambda_{\rm b} < V_0/8$, i.e. when this condition is satisfied, 
inflation ends in the low energy limit. When $n \neq 2$, we obtain,
at the end of inflation,
\begin{equation}
V_{\rm end} = V_0 
\left( \frac{2 n^2 \lambda_{\rm b}}{V_0} \right)^{n/(n+2)} \,,
\end{equation}
and $N$ $e$--foldings prior to the end,
\begin{equation}
\label{VN2}
V_N = V_{\rm end}~ \left[ n+(n+2)N \right]^{n/(n+2)} \,.
\end{equation}

The brane tension is no longer independent of the mass scale $V_0$ but 
is related to it by,
\begin{equation}
\lambda_{\rm b} = V_0\left( \frac{\kappa^4}{600 \pi^2}
\frac{V_0}{n^2}\frac{1}{A_{\rm S}^2} \right)^{\frac{n+2}{4-n}}
\left[ 2n \left( n+(n+2)N \right) \right]^{\frac{4n+2}{4-n}} \,.
\end{equation}
when $n \neq 4$, otherwise $\lambda_{\rm b}$ is completely free and only
$V_0$ is fixed, yielding
\begin{equation}
V_0 = \frac{75 \pi^2}{256 ~\kappa^4} ~\frac{1}{(1+N)^3} A_{\rm S}^2 \,.
\end{equation}

For the scalar and tensor spectral indices and their derivatives we find,
\begin{eqnarray}
\label{spectral2}
n_{\rm S} -1 &=&  - \frac{2+4n}{n+(2+n)N} \,, \\
n_{\rm T} &=& - \frac{3n}{n+(2+n)N} \,, \\
\frac{d n_{\rm S}}{d \ln k} &=&
- 2 ~\frac{2 + 5 n + 2 n^2}{\left(n+(2+n)N\right)^2} \,, \\
\frac{d n_{\rm T}}{d \ln k} &=&
- 3~\frac{2n + n^2}{\left(n+(2+n)N\right)^2} \,.
\end{eqnarray}
which reduce to Eqs.~(\ref{spectral1}) in the limit $n \rightarrow \infty$.

For negative powers of $n$, 
the precise number of $e$--foldings corresponding to 
the scale entering the Hubble radius today is easily determined from
Eqs.~(\ref{efolds1}) and (\ref{VN2}).

\begin{figure}[ht]
\includegraphics[width=8.5cm]{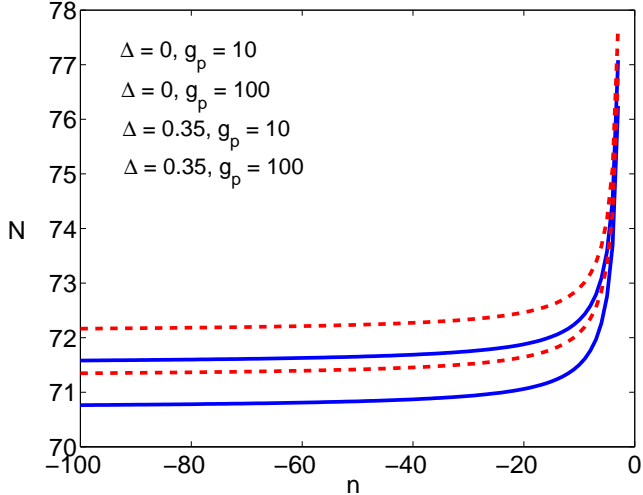}
\caption{\label{findN} Dependence of the number of $e$--foldings
on $n$. From top to bottom, for $\Delta = 0$, $g_{\rm p} = 10$;
$\Delta = 0$, $g_{\rm p} = 100$;
$\Delta = 0.35$, $g_{\rm p} = 10$;
$\Delta = 0.35$, $g_{\rm p} = 100$. }
\end{figure}

In Fig.~\ref{findN} we show the dependence of the number of $e$--foldings
$N$ with $n$ for four different combinations of the time delay between
the production of particles and the end of inflation and the number of
fields experiencing particle production.

For completeness we review the main features of power law potentials in
the case of standard inflation. In this case, 
there is no freedom for the
mass scale $V_0$ which is fixed by the density perturbations as
\begin{equation}
V_0 = \left[ \frac{\kappa^4}{75 \pi^2}\frac{1}{n^2}
\left( 2n (n/4+N) \right)^{\frac{n+2}{2}} \right]^{-1} A_{\rm S}^2 \,.
\end{equation}

The corresponding spectral indices are 
\begin{eqnarray}
n_{\rm S} -1 &=& (n/4+N)~\frac{d n_{\rm S}}{d \ln k} = 
-\frac{1}{2}~\frac{n+2}{n/4+N} \,,
\\
n_{\rm T} &=& (n/4+N)~\frac{d n_{\rm T}}{d \ln k} =
-\frac{1}{2}~\frac{n}{n/4+N} \,.
\end{eqnarray}
%
\subsection{Future observational constraints}
\begin{figure}[ht]
\includegraphics[width=8.5cm,height=12cm]{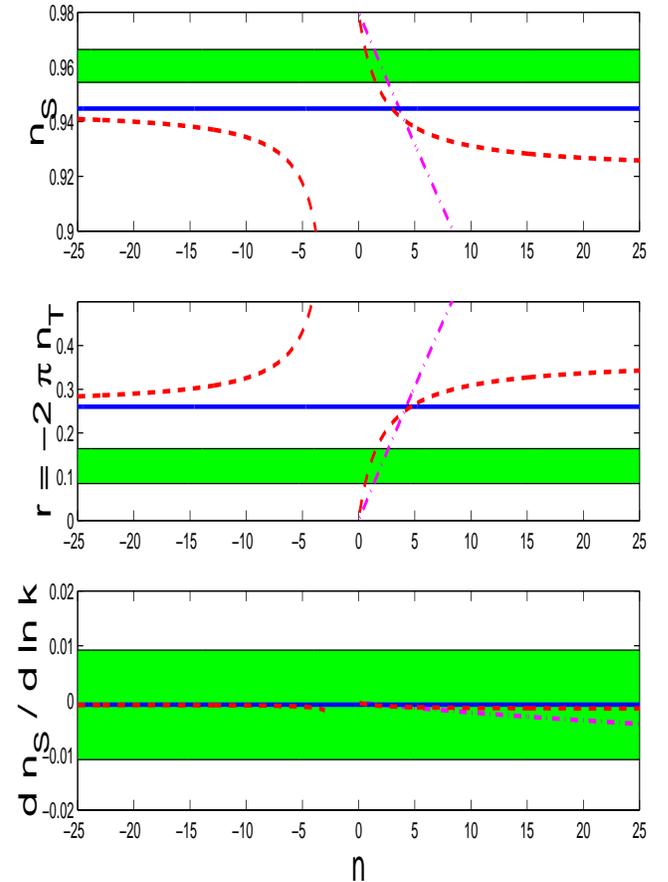}
\caption{\label{bolas} Dependence of the spectral index, the
ratio of tensor to scalar perturbations and the derivative of the
spectral index on the
parameter $n$. The solid line: exponential potential; dashed line:
power law potential in brane worlds; dash dotted line: power law in
standard inflation. In the case of the exponential potential and
power law with negative $n$ we used Eqs.~(\ref{efolds1}), (\ref{VN})
and (\ref{VN2}) with $g_{\rm p} = 100$ and $\Delta = 0.35$, to define
$N$. For the power law cases with $n$ positive we used $N = 50$.}
\end{figure}

In Fig.~\ref{bolas} we show how a selection of different
observational parameters depend on the precise value of $n$ 
comparing standard inflation with brane world inflation. 
In particular,
we consider here the spectral index $n_{\rm S}$, the ratio of tensor
to scalar perturbations 
$r \equiv 4 \pi A_{\rm T}^2/A_{\rm S}^2 = - 2 \pi n_{\rm T}$ and the
derivative of the spectral index 
$d n_{\rm S}/ d \ln k$. 
For the positive power law case, reheating can happen through conventional
mechanisms, and hence we take the number of $e$--foldings corresponding to 
the scale entering the Hubble radius today of $N= 50$.
The filled areas correspond to
estimated error bars on these
observables based on upcoming Planck observations,
according to Ref.\cite{grivell}. We quote, 
$\delta n_{\rm S} = 0.006$, $\delta r = 0.04$ and 
$\delta d n_{\rm S}/ d \ln k = 0.01$. The error bars are centred on a
$\phi^2$ standard inflationary cosmology for the sake of illustration. There
are several pieces of information one can extract from this figure.
Firstly, a $\phi^2$ theory in brane worlds is degenerate with the same
theory in standard inflation. The case becomes worse for a $\phi^4$
theory as the exponential potential presents similar values for all the
observables. Secondly, in brane worlds, an exponential and power law
with $n < -10$ are degenerate potentials. Thirdly, if we could rule
out the exponential potential, the remaining models would be truly
non--degenerate only if $|n| > 20$. And finally, it appears that the 
derivative of the
spectral index will be unable to give us any information about the
underlying inflationary cosmology. In summary, based on this small 
sample, it appears that 
upcoming satellite experiments wont be able to 
{\it conclusively} discriminate between the various models of inflation. 

In the next section, we go one step further trying to use the same
field to link the
early inflation epoch with the current accelerated expansion of the
universe, which is strongly
suggested by supernovae type Ia (SN Ia) observations.

\section{Quintessence}
In the case of reheating by means of gravitational production of
relativistic particles, the scalar field does not totally decay. Many authors
\cite{PV,quinfl}
have suggested that the field could then dominate the energy 
density at late times,
providing the dark energy that appears to be driving the 
observed acceleration of the universe today 
\cite{data}.

In the brane world picture the potential does not have to have a
shallow slope to satisfy the slow--roll conditions followed by a steep
fall where the latter break, as in standard
inflation. In fact, we have seen that the potential can be exponentially
steep and still inflate at early times and at the point where the
brane correction becomes negligible inflation comes gracefully to an
end. This feature broadens  the
range of potentials one can consider for ``quintessential inflation''
provided the brane effects become negligible before nucleosynthesis 
\cite{huey,majumdar,sahni}.
 
Using the modified Friedmann equation 
(\ref{Friedmannmodify}), Huey and Lidsey\cite{huey} showed that
for a pure inverse power law potential,
the requirement that the field was both sub--dominant prior to
nucleosynthesis, and tracked the matter and radiation in the 
universe before today
implied that the value of the power $n$ in Eq.~(\ref{powerlawpot}), 
is so large (and negative) that the corresponding equation of
state ($w_{\phi} \equiv p_{\phi}/\rho_{\phi}$) is too large to 
give an accelerating universe at this epoch. They concluded that 
within the class of inverse power law potentials, the only successful model 
would correspond to a kind of ``creeping quintessence'', 
in which the energy density of
the field is kinetic energy dominated from its value at the end of
inflation to its present value. The field effectively remains 
frozen (with equation of state $w_{\phi} = -1$) until the
energy density of matter 
matches its energy density, which occurs only around a redshift 
$z \approx 1$. 
At this point the field starts rolling down its potential again,
and dominates the total energy density
slowly raising the equation of state from $-1$, thereby explaining the
present day acceleration of the universe. 
Although an interesting scenario, its similarity
to a pure cosmological constant at late times, makes the evolution of this 
quintessence field very difficult to confirm on the basis of upcoming 
SN Ia or CMB observations, 
especially if the equation of state of dark
energy is very close to $-1$ as observations suggest \cite{pier}.
However, if the contribution of the
quintessence field to the total
energy density is significant, it can increase the expansion rate
resulting in a shift in the positions and an increase in the amplitude
of the peaks of the angular power spectrum \cite{powerspec}, hence
opening up the possibility that we may 
be able to discriminate a quintessence field from a bare cosmological 
constant type behaviour.

In what follows we analyse a class of scalar
potentials in which the energy density of
the quintessence field tracks the background energy density during
the epoch of decoupling. We are able to constrain the parameters of
the potentials by employing the values allowed on the  
fractional density of a scalar field, both at the epoch of nucleosynthesis
and  at the time of recombination \cite{bean}:
\begin{eqnarray}
\label{constraints}
\Omega_{\phi}^{\rm nuc} < 0.045 ~, \hspace{1cm} 
\Omega_{\phi}^{\rm rec} < 0.39~.
\end{eqnarray}

In the next subsections we will consider the
two possible cases of evolution of the scalar field between the end of
inflation and the epoch the attractor is reached.

\subsection{Monotonic evolution}
After the initial period of inflation and following that of the 
kinetic energy domination, the scalar
field is known to freeze at a value given by \cite{zlatev}
\begin{equation}
 \phi_{\rm ke} = \phi_{\rm end} + \frac{\sqrt{6}}{\kappa}
\left[ 1 + \frac{1}{2} \ln \left( \rho_{\phi}^{\rm end}/
                                       \rho_{\rm R}^{\rm end} \right)
\right] \,.
\end{equation}

\begin{figure}[ht]
\includegraphics[width=8.5cm]{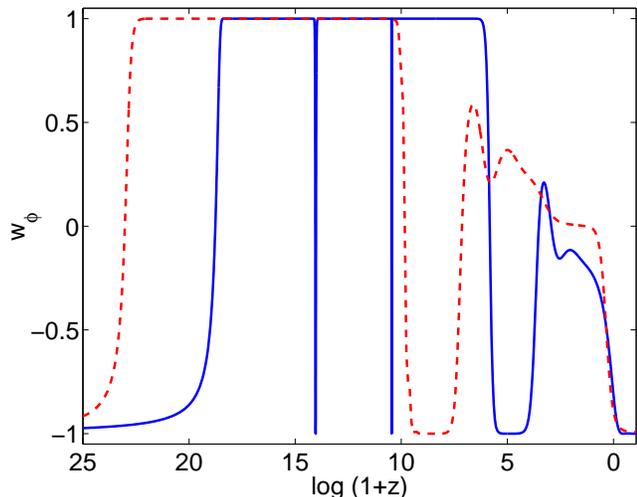}
\caption{\label{eqst} The evolution of the equation of state of 
the scalar field after inflation. 
The dashed line represents the 2EXP model with 
$\alpha = 4$, $\beta = 0.1$ and the solid line the SUGRA model with
$n = -23$.}
\end{figure}

Using Eq.~(\ref{ratio}) and the results of the previous section 
for an exponential scalar potential, we can calculate the bounds on
the allowed slope of the potential.
We require the field to enter the 
scaling regime (whereupon $\Omega_{\phi} = 4/\alpha^2$ when radiation is
dominant) after nucleosynthesis but before 
radiation--matter equality, i.e.,
\begin{equation}
4 \rho^{\rm eq}/\alpha^2 \lesssim V(\phi_{\rm ke}) \lesssim 4 \rho^{\rm
nuc}/\alpha^2 \,.
\end{equation}
These requirements impose the following bounds:
$3.6 \lesssim \alpha  \lesssim 4.8$ for 
$g_{\rm p} = 100$, $\Delta = 0.35$ and 
$3.4 \lesssim \alpha  \lesssim 4.5$ for 
$g_{\rm p} = 10$, $\Delta = 0$,
results found to be 
in good agreement with our numerical integrations (see row
corresponding
to zero oscillations in Table \ref{tabela}).
Now, in the standard Friedmann cosmology, a pure exponential potential
with a slope within this range of parameters has a stable
attractor solution for the energy density that scales 
with the background \cite{scaling}.
 Therefore, this solution does not provide a late time 
accelerating universe.
A number of modifications to this type of potential have been proposed and 
investigated in depth which do fit the observational data 
\cite{albrecht,nelson1,sahni1}. These models present a pure exponential term
dominating for almost the entire history of the universe and only
recently the other features of the potentials become dominant to
provide the accelerating universe, hence, the above constraints also
apply to the exponential part of these potentials. In Fig.~\ref{eqst},
we show the evolution of the equation of state
 of the scalar field for the two
exponentials potential (2EXP) \cite{nelson1,majumdar}
\begin{equation}
V(\phi) = V_0 \left( e^{\alpha \kappa \phi} + e^{\beta \kappa \phi} 
\right) \,.
\end{equation}
We would like to make two remarks on the pure exponential case. 
The first concerns
the small range of the parameter $\alpha$ where this scenario can be
implemented revealing some degree of  fine tuning. 
The second, refers to the smallness of the parameter itself.
In this range of parameters, the contribution of the field is large
enough for the integrated Sachs--Wolf (ISW) effect to increase the power on
large scales. Under the assumption that the anisotropies we see today are
primordial, the normalisation of the power spectrum to COBE at
the multipole 
$\ell = 10$ therefore, results in a wrong suppression of
the small scales anisotropies. Hence, for these type of potentials, 
the analysis of the angular
power spectrum must be made with special care to take into account
the ISW effect.

\subsection{Oscillatory evolution}
In the case of very steep potentials, the kinetic energy of the field
at the end of inflation can be large enough to overtake the potential 
by a huge factor. 
Consequently, the field might freeze at
a value corresponding to an energy density lower than the energy
density of the background at equality, or even lower than today's
energy density. 
One then requires an additional mechanism to slow down 
the field's evolution. Potentials with  a minimum can provide the
necessary solution.

\begin{figure}[ht]
\includegraphics[width=8.5cm]{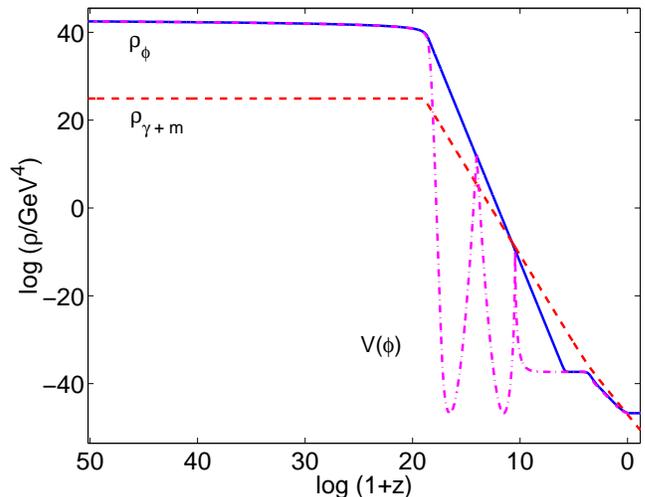}
\caption{\label{rhos} The evolution of the energy densities of
the scalar field (solid) and of the background fluid (dashed) 
and the evolution of the
scalar potential (dash--dotted) for the SUGRA model with $n = -23$.}
\end{figure}

For potentials with a minimum,
the field can perform several wide damped oscillations around
its minimum while energy is continually being converted between
kinetic and potential. Hence, extra time for the friction term in
Eq.~(\ref{eqom}) to slow down the field.
In the regime of wide oscillations the equation of state of the field 
is close to unity
as the potential is very steep and 
looks like $V \propto \phi^n$ for $n \gg 1$.
Hence, in practice, the field energy density continues decaying as
$a^{-6}$.
In Fig.~\ref{eqst} we show the evolution of the equation of 
state of the field for a potential with a minimum (solid
line). Fig.~\ref{rhos} shows the evolution of the energy densities of
the scalar field and of the background fluid and the evolution of the
scalar potential for the same potential with a minimum.
After inflation
the field quickly becomes kinetic energy dominated. The sharp peaks
with $w_{\phi} = -1$ in Fig.~\ref{eqst} or the epochs when $V(\phi) =
V_{\rm minimum}$ in Fig.~\ref{rhos}
represent the points at which the field passes through the minimum. In this
example, the field becomes sub-dominant just before nucleosynthesis as
$\Omega_{\phi}$ drops below 0.5 before $z > 10^{10}$.      
After a few oscillations the field has slowed down enough
and freezes (for $z \approx 10^5$, in this example). 
From here the usual dynamics resume. When the background
fluid's energy density drops, the field reaches the attractor
solution. The contribution of the field increases when it approaches
the minimum of the potential, then the equation of state quickly
decreases to $-1$ accelerating the universe by today.

We give two
examples corresponding to models presented before in the literature.
The 2EXP potential
with $\alpha/\beta < 0$, satisfies the above requirements.
For the particular case of symmetric slopes i.e. $\beta = -\alpha$
\cite{rubano}, we
find that to satisfy the nucleosynthesis bound  and for
the field to scale after this period and no later than 
radiation--matter equality, the slopes must satisfy the constraints
presented in Table \ref{tabela}.
\begin{table}[ht]
\begin{tabular}{c|c|c}
{\bf oscillations }& {\bf $g_{\rm p} = 100$, $\Delta = 0.35$ ~~} &
{\bf $g_{\rm p} = 10$, $\Delta = 0$ ~~} \\ \hline 
0 & $ 3.6 \lesssim |\alpha| \lesssim 4.8$ &
    $ 3.1 \lesssim |\alpha| \lesssim 4.1 $      \\
$1^-$ & $ 5.9 \lesssim |\alpha| \lesssim 7.0 $ &
    $ 5.1 \lesssim |\alpha| \lesssim 6.1$      \\
$1^+$ & $ 10.0 \lesssim |\alpha| \lesssim 11.6 $ &  
    $ 7.9 \lesssim |\alpha| \lesssim 9.4$    \\
2 & $ 12.9 \lesssim |\alpha| \lesssim 19.0 $ & 
      $ 10.7 \lesssim |\alpha| \lesssim 15.3 $   \\
3 & $ 20.3 \lesssim |\alpha| \lesssim 26.5 $ & 
      $ 16.6 \lesssim |\alpha| \lesssim 21.3 $  \\
\end{tabular}
\caption{\label{tabela} Constraints on tracking quintessential 
inflation for 2EXP with symmetric slopes. $1^-$ refers to the case in
which the field freezes when rolling up the potential and $1^+$ to the
case in which the field is rolling down before it stops.}
\end{table}

The second example is for the SUGRA type potentials \cite{sugra},
\begin{equation}
V(\phi) = V_0~(\kappa \phi)^n~
\exp \left[ \frac{1}{2} (\kappa \phi)^2 \right] \,.
\end{equation}
For initial conditions of $\phi \ll \phi_{\rm minimum}$, 
the bounds in Eq.~(\ref{constraints}) above place
a restriction of $ -78 \lesssim n \lesssim -23$ for 
$g_{\rm p} = 100$, $\Delta = 0.35$ and 
$n \lesssim -86$ for 
$g_{\rm p} = 10$, $\Delta = 0$, corresponding to two
oscillations of the field (see Fig.~\ref{eqst}) and 
$w_{\phi} \approx -0.8$, today. 

To obtain the quoted results, the initial conditions of the scalar field
in our simulations 
were such that they gave a number of $e$--foldings of inflation
consistent with the number of $e$--foldings corresponding
to scales entering the Hubble radius today in Eq.~(\ref{efolds1}). 
We have taken $\Omega_{\phi} \approx 0.7$, today.

\section{Conclusions}
We have demonstrated, giving examples, of a number of classes of scalar
potentials that provide realistic models of ``tracking quintessential
inflation''. We have seen that the parameter space of potentials
described by a pure exponential at early times is highly contrived
on the basis of nucleosynthesis bounds and under the requirement that the
scalar field had reached the attractor by radiation--matter equality.
However, potentials with a minimum offer a larger range of parameter
values for which these requirements are satisfied. 

For the potentials we have considered, successful realisation of 
``quintessential inflation'' requires potentials with very steep slopes
in order to satisfy the nucleosynthesis bound, therefore we expect for all
of these models similar values for the observables, 
i.e. $n_{\rm S} \approx 0.94$
and $r \approx 0.27$ (using Eqs.~(\ref{spectral2}) for large $n$ 
with $N \approx 70$ or from Fig.~\ref{bolas}). 

We have seen in Sec.~\ref{braneinflation} and Fig.~\ref{bolas} that
these models can be degenerate to a $\phi^4$ theory either in standard
or in brane world inflation. It is important to know then, if the type
of potentials considered here fit both the reconstruction of the 
quintessence potential from SN Ia observations
\cite{reconstsne} and the
reconstruction of the inflaton potential from CMB observations 
\cite {reconstcmb} giving indications of a unified theory of ``quintessential
inflation''.

\section*{Acknowledgements}
We acknowledge conversations with Dominic Clancy and Andrew Liddle. We
thank Pier Corasaniti and David Wands for helpful comments on the
manuscript. N.J.N. is supported by Funda\c{c}\~{a}o para a Ci\^{e}ncia e a
Tecnologia (Portugal).


\end{document}